\documentclass[prd,preprint,aps,amsfonts,amssymb,nofootinbib,tightenlines]{revtex4}
\usepackage{amscd}
\usepackage{amsmath}
\usepackage{bm}
\usepackage[dvips]{color}                 %%% for colours in the graphs
\usepackage{graphicx}
\usepackage{eufrak}
\usepackage{eucal}

\newcommand{\beq}{\begin{equation}}
\newcommand{\eeq}{\end{equation}}
\newcommand{\bea}{\begin{eqnarray}}
\newcommand{\eea}{\end{eqnarray}}

\newcommand{\benn}{\begin{displaymath}}
\newcommand{\eenn}{\end{displaymath}}
\begin{document}

\title{\bf  \large{Surface Energy in Cold Asymmetrical Fermion Superfluids}}

\author{Heron Caldas\footnote{{\tt hcaldas@ufsj.edu.br}} }
\affiliation{Universidade Federal de S\~{a}o Jo\~{a}o del Rey, S\~{a}o Jo\~{a}o del Rei, 36300-000, MG, Brazil}

\begin{abstract}
We derive the energy of the surface between the normal and superfluid components of a mixed phase of a system composed of two particle species with different densities. The surface energy is obtained by the integration of the free energy density in the interface between the two phases. We show that the mixed phase remains as the favored ground state over the gapless phase in weak coupling. We find that the surface energy effects emerge only at strong coupling.

\end{abstract}
\maketitle
\bigskip

The issue of superfluidity of fermionic atoms in cold atomic (laser) traps has motivated a wide community as a consequence of the improvement of experimental techniques for dealing with these systems in the laboratory~\cite{Kinast,Bartenstein,Regal,Zwierlein1,Chin,Zwierlein2,Partridge}. A richer physics emerges when the populations of the two candidates to form pairing are unequal. There will be unpaired fermions in the system. In this fascinating scenario, various alternatives for the ground state of these asymmetrical fermionic systems have been proposed in the literature~\cite{Sarma,loff,internalgap,Paulo,Heron,Forbes,Sedrakian}. 

In Refs.~\cite{Paulo,Heron} the energies of the mixed phase (MP) and the breached-pair superfluid phase~\cite{internalgap} were compared for different density asymmetries of a fermionic gas composed of two particle species. It was shown that phase separation is the energetically favored ground state. This figure appears to gain strong experimental support~\cite{Ketterle,Hulet}. 

In these calculations the energy of the surface between the two equilibrium phases of the MP has been neglected, since this term is negligible in the thermodynamic limit (infinite volume), as considered in~\cite{Paulo,Heron}. In principle, the cost associated with the interface separating the phases could play an important role in finite atomic systems. For instance, the surface tension between the normal and superfluid phases can result in deformations of the minority component that are quit similar to those observed by the Rice experiments~\cite{Hulet}. However, the very poor knowledge of the surface energy function properties precludes a fully quantitative description of the phase separation stability in asymmetrical fermionic systems. 

In this paper we derive the surface energy that could be present in ultracold asymmetrical fermion superfluids, in order to verify the stability of the inhomogeneous phase separation~\cite{Paulo,Heron} over the homogeneous gapless superfluid (HGS) phase~\cite{internalgap,Forbes}.

\subsubsection{The Mixed Phase Total Energy}

We consider a dilute fermionic gas made up of $a$ and $b$ particle species (that could represent, for instance, two hyperfine-spin states of fermionic atoms, such as $^{6}{\rm Li}$ or $^{40}{\rm K}$) with densities $n_a$ and $n_b$, in a ``box" of volume $V_{box}$. The MP state is formed by $\tilde n_a$ and $\tilde n_b$ (unpaired) atoms in the normal phase, immersed in a sea of $n_a^{\rm{BCS}}=n_b^{\rm{BCS}}=n$ (paired) atoms in the superfluid phase. Thus, the number densities $n_a$ and $n_b$ of the two atom species are accommodated within the box as $n_{a,b}=x \tilde n_{a,b} + (1-x)n$. The parameters $x$ and $n$ are found by minimization of the energy of the  mixed (inhomogeneous) phase~\cite{Paulo,Heron} which will be defined below.

For practical calculations we consider that the normal asymmetric state is in a bubble of radius $R$ and the symmetric superfluid state is in the rest of the volume, defined by a Wigner-Seitz cell of radius $R_W$~\cite{Reddy}. Then the total energy of the MP is given by:

\bea
\label{eq1_1}
&&{\rm E}^{\rm MIX}(n_a,n_b,\sigma)\\
\nonumber
&=&V_{bu} {\rm E}^{\rm N}(\tilde n_a,\tilde n_b)+(V_{box}-V_{bu}){\rm E}^{\rm BCS}(n) + 4 \pi R^2 \sigma \\
\nonumber
&=&V_{box} \left[ x{\rm E}^{\rm N}(\tilde n_a,\tilde n_b)+ (1-x){\rm E}^{\rm BCS}( n) + \frac{3}{R}  x \sigma \right],
\eea
where $V_{bu}=\frac{4}{3} \pi R^3$, $x=V_{bu}/V_{box}$ is the volume fraction, $\sigma$ is the surface energy density, to be calculated below, ${\rm E}^{\rm N}(\tilde n_a,\tilde n_b)=\frac{(6\pi^2)^{\frac{5}{3}}}{20\pi^2}
\left[ \frac{\tilde n_a^{5/3}}{m_a}+ \frac{\tilde n_b^{5/3}}{m_b} \right]$ is the energy of the normal (unpaired) particles, and ${\rm E}^{\rm BCS}(n)=\frac{(6 \pi^2 n)^{5/3}}{20 \pi^2 M}-\frac{M}{2 \pi^2} (6 \pi^2 n)^{1/3} \Delta_0^2$ is the energy of the superfluid state, where $\Delta_0$ is the mean field BCS gap parameter in the weak coupling approximation:

\beq
\label{eq5}
\Delta_0  = \frac{4 \mu^{\rm BCS}}{e^2} e^{\frac{-\pi}{2 k_F |a|}},
\eeq
where $\mu^{\rm BCS}=\mu_a^{\rm BCS} + \mu_b^{\rm BCS}$, $k_F=\sqrt{2 M \mu^{\rm BCS}}$ is the Fermi surface of the asymmetrical BCS superfluid, $M=\frac{m_a m_b}{m_a + m_b}$ is the reduced mass, and $a$ is the two body scattering length. As we pointed out earlier, for sufficiently large systems, meaning infinite volume, the last term in Eq.~(\ref{eq1_1}) ($\propto R^2$) is negligible in comparison with the first two ones ($\propto R^3$).

The limits found for the volume fraction are $x=1$ if $n_b >> n_a$ (the whole system is in the normal phase), and $x=0$ if $n_b=n_a$ (the whole system is in the BCS phase)~\cite{Paulo,Heron}. Since in both limits the surface energy must be absent, we write the MP energy density ${\rm E}^{\rm MIX}(n_a,n_b,\sigma)/V_{box}$ as

\bea
\label{eq1_2}
&&\bar{{\rm E}}^{\rm MIX}(n_a,n_b,\sigma)={\rm E}^{\rm MIX}(n_a,n_b)+{\rm E}^{\rm SU}(\sigma)\\
\nonumber
&=& \overbrace{x {\rm E}^{\rm N}(\tilde n_a,\tilde n_b)+(1-x){\rm E}^{\rm BCS}(n)}^{{\rm E}^{\rm MIX}(n_a,n_b)}\\
\nonumber  
&+& \overbrace{ \frac{3}{R} \sigma \left[ x \Theta(1/2-x)+(1-x)\Theta(x-1/2) \right] }^{{\rm E}^{\rm SU}(\sigma)},
\eea
where $\Theta (x)$ is the Heaviside ``step-function'', defined as $\Theta (x)=1$ if $x>0$ and $\Theta (x)=0$ if $x<0$, ${\rm E}^{\rm MIX}(n_a,n_b)$ is the energy density of the normal and superfluid phases and ${\rm E}^{\rm SU}(\sigma)$ is the surface energy density. The lowest MP energy $\bar{{\rm E}}^{\rm MIX}(n_a,n_b,\sigma)$ is found by minimizing ${\rm E}^{\rm MIX}(n_a,n_b)$ with respect to $n$ and $x$, and the obtained $x_{min}$ is plugged in ${\rm E}^{\rm SU}(\sigma)$ above.

\subsubsection{The Calculation of the Surface Energy}

We will now compute the surface energy between the normal and superfluid phases. The derivation is similar to the work of Reddy and Rupak~\cite{Reddy} which calculated, in the quark matter context, the surface tension between the normal and superconducting phases. To achieve this, they included the leading order gradient contribution to the free energy, whose coefficient (termed there $\kappa_{2SC}^{(2)}$) was approximated as being independent of the chemical potential asymmetry and spatial variations of $\Delta$. In Ref.~\cite{Reddy} $\kappa_{2SC}^{(2)}$ was determined by matching the mass of the fourth and eight gluon in the effective theory to the microscopic calculation while, here, we calculate the coefficients within the model under consideration.

To describe the microscopic theory at the interface between the normal and superfluid phases, we begin writing the partition function:

\beq
\label{gf1}
Z=\int D[\psi_{a,b}] D[\psi_{a,b}^\dagger] e^{[- S(\psi, \psi^\dagger)]},
\eeq
where $S(\psi, \psi^\dagger)= \int d \tau \int dx~ {\cal L}$, with ${\cal L}$ being the BCS Lagrangian

\bea
\label{BCS1}
{\cal L}= \sum_{i=a,b} \psi_i^\dagger(x) \left( \partial_{\tau} - \frac{\bigtriangledown^2}{2 m_i} -\mu_i \right) \psi_i(x)- g \psi_{a}^\dagger(x) \psi_{b}^\dagger(x) \psi_{a}(x) \psi_{b}(x),
\eea
where $\psi_i(x)$ describes a fermionic atom of the type $i=a, b$ at $x=(x, \tau)$, $g >0$ is the coupling constant, and $\mu_{a,b}$ is the chemical potential of the non-interacting species. After equilibrium is reached the particles will have rearranged chemical potentials in each component of the MP. For this dilute system ($\delta \mu=\frac{\mu_b-\mu_a}{2}<\mu_{a,b}, \rm{and~especially}~\delta\mu << \Delta_0)$, we consider that the local density approximation (LDA) holds, i.e. the chemical potentials in the individual phases of the MP are homogeneous and could be regarded as effective, including the contribution from the (external) trapping potential. We notice, however, that it appears that the LDA does not hold in the Rice experiments~\cite{Hulet}, and is consistent with the results of the MIT experiments~\cite{Ketterle,Ketterle2}.

We integrate out the fermions after introducing the usual Hubbard-Stratonovich field $\Delta(\tau, x)$, to obtain $Z= \int D \Delta D \Delta^*  exp(- S_{eff}[\Delta, \Delta^*])$. The effective action is expressed by

\beq
\label{EffA}
S_{eff}[\Delta, \Delta^*]= \int \frac{|\Delta(x)|^2}{g} - {\rm Tr} \ln G^{-1}[\Delta(x)],
\eeq
where $\int=\int_0^{\beta} d \tau \int dx$, and $G^{-1}[\Delta(x)]$ is the inverse of the Nambu propagator:

\beq
 G^{-1}[\Delta(x)] = \left[
\begin{matrix}  \partial_{\tau} - \frac{\bigtriangledown^2}{2 m_b} -\mu_b & \Delta(x)\\
  \Delta^*(x) & \partial_{\tau} + \frac{\bigtriangledown^2}{2 m_a} + \mu_a
\end{matrix} \right].             
\eeq

Near the critical temperature $T_c$, where $\Delta/{T K_B}$ can be considered a small quantity, the Ginzburg-Landau (GL) theory provides a successful description of the phase transition~\cite{GL}. Although we are considering a zero temperature system, we have a physics motivated approximation. Since the gap varies from $\Delta_0$ (on the superfluid side) to $0$ (on the normal side) at the interface between these two phases, we consider a slowly varying $\Delta$ and expand the action (in momentum-frequency space) up to forth order in $\Delta(q)$. At T=0 one should, in principle, include all terms of the expansion. However, if the spatial variations of $\Delta(x)$ occur on a length scale that is numerically large compared to $\Delta_0^{-1}$, it is a good approximation to keep only the leading term of the gradient in the expansion. This condition is guaranteed if $\nabla(\Delta(x))/\Delta_0^2<<1$~\cite{Reddy}, and we have verified that this is satisfied in the present work. This approach has been explicitly taken in Refs.~\cite{Reddy,Mueller}, where the solution found for $\Delta(x)$ justifies the approximation.

\beq
\label{EffA2}
S_{eff}[\Delta, \Delta^*]= \sum_{q} [{\cal A} |\Delta(q)|^2 + \frac{\beta}{2} |\Delta(q)|^4],
\eeq
where $q=(\vec{q},i\omega_{m})$, with $\omega_{m}=2m \pi T$, ${\cal A}=\left(\frac{1}{g}-T \sum_{n} \int \frac{d^3k}{(2 \pi)^3}\frac{1}{[-i \omega_n +i \omega_m - \epsilon_{-\vec{k}+\vec{q}}^a][i \omega_n - \epsilon_{\vec{k}}^b]} \right)\equiv \left( \frac{1}{g}-\Pi(\vec{q},i\omega_{m}) \right)$, $\omega_n=(2n+1)\pi T$, $ \epsilon_{-\vec{k}+\vec{q}}^a=\frac{(-\vec{k}+\vec{q})^2}{2 m_a}-\mu_a$, $ \epsilon_{\vec{k}}^b=\frac{\vec{k}^2}{2 m_b}-\mu_b$. As we are interested in the static limit of the corrections, we shall set $i \omega_m=0$. We also set $m_a=m_b=m$. This brings about an enormous simplification to the problem without compromising the asymmetry of the system. The same consideration has been made in Ref.~\cite{Carlson}. The coupling $g$ can be related to the (s-wave) scattering length $a$ by $\frac{1}{g}=\frac{m}{4 \pi |a|} + \int\frac{d^3 k}{(2 \pi)^3} \frac{1}{2 \xi_k}$, where $\xi_k=\frac{\vec{k}^2}{2m}$~\cite{Melo}. Then we use this expression to regulate the ultraviolet divergence from the vacuum part of ${\cal A}$. After performing the Matsubara frequency summation, the matter part of $\Pi(\vec{q},0)$ is written as $\Pi^{m}(\vec{q},0)=\int \frac{d^3k}{(2 \pi)^3}\frac{n_f(\epsilon_{\vec{k}+\vec{q}/2}^a)+ n_f(\epsilon_{\vec{k}-\vec{q}/2}^b)}{\epsilon_{\vec{k}+\vec{q}/2}^a+\epsilon_{\vec{k}-\vec{q}/2}^b}$, where $n_f(x)\equiv1/(exp(x/T)+1)$ is the Fermi distribution function. Expanding $n_f(\epsilon_{\vec{k}\pm \vec{q}/2}^{a,b})$ for $|\vec{q}|<< k_F^{a,b}$, with $k_F^{a,b}=\sqrt{2 m \mu_{a,b}}$, and taking the zero temperature limit we obtain

\bea
\label{EffA4}
\Pi^{m}(\vec{q},0)=\int \frac{d^3k}{(2 \pi)^3} \frac{\Theta^{a}-\frac{\delta^{a}}{2 k_F^{a}} (q^2/4-kq\cos\theta) + (a\leftrightarrow b)}{1/m\left[k^2+q^2/4-m \mu \right]},
\eea
where $\Theta^a \equiv \Theta (k_F^{a}-|k|)$, $\delta^a \equiv \delta(|k|-k_F^{a})$ is the delta function, and $\mu=\mu_a+\mu_b$. The evaluation of the integral in the equation above yields ${\cal A}(|\vec{q}|)=\alpha+\gamma|\vec{q}|^2 + 
{\cal O}\left( \frac{|\vec{q}|}{k_F^{a,b}} \right)$, where $\alpha=N_0 \left( \frac{\pi}{4 \bar{k}_F |a|} - 1 \right)$, and $\gamma=\frac{m}{(4 \pi)^2 \bar{k}_F}$. Here we have defined, $N_0=\frac{ m \bar{k}_F}{\pi^2}$, and $\bar{k}_F=\frac{k_F^a+k_F^b}{2}$ is the ``average'' Fermi surface. This will allow us to study which phase is the most likely ground state as a function of the non dimensional interaction parameter $1/\bar{k}_F|a|$.

We now obtain the equation $\frac{\delta~S_{eff}}{\delta \Delta(q)^*}=0$, and then take its Fourier transform to $(x,0)$:

\beq
\label{gl}
[\alpha + \beta |\Delta(x)|^2 - \gamma \nabla^2] \Delta(x)=0.
\eeq
The solutions of the equation above depend on $\alpha$, and can be divided into the limits $a<0$ and $a>0$.

\begin{center}
A. $\mu>0$, $a<0$
\end{center}
In absence of gradients, we obtain a space-independent solution that corresponds to the gap inside the superfluid region:
\beq
\label{si}
|\Delta|^2=-\frac{\alpha}{ \beta},
\eeq
which has meaning only if the ratio $\frac{\alpha}{ \beta}$ is negative. The trivial solution refers, obviously, to the gap inside the normal region.

\begin{figure}[t]
\includegraphics[height=2.8in]{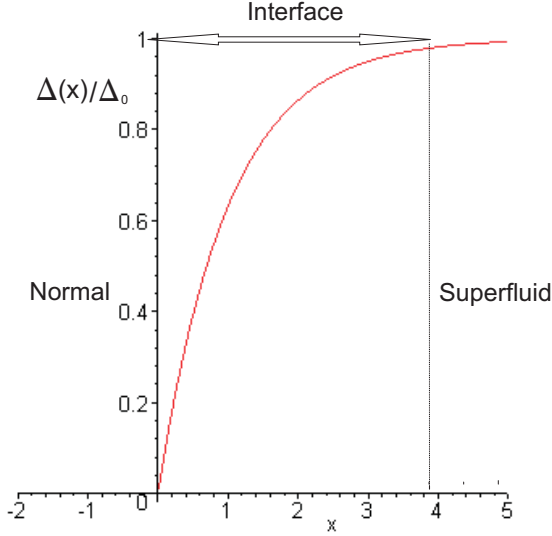}
\caption{\label{gapprofile}\textit{Surface region between normal and superfluid components of the mixed phase. } }
\end{figure}

Seeking now space-dependent (one-dimensional) solutions, let us define a real function $\Delta(x)=A f(x)$~\cite{Fetter}, where $A=\left(|\frac{\alpha}{\beta}| \right)^{1/2}$. Then Eq.~(\ref{gl}) can be written for $\alpha < 0$ (as will be clear in the expression for the energy below, a stable theory needs $\beta>0$) as

\beq
\label{sd}
\xi^2 \frac{d^2}{dx^2}f(x) + f(x) -f(x)^3=0,
\eeq
where $\xi^2=\frac{\gamma}{|\alpha|}$. We find the solution of Eq.~(\ref{sd})~\footnote{Eq.~(\ref{sd}) also admits the solution $f(x)= \tanh(x/\sqrt{2} \xi)\theta(x)$, as well as the ansatz $f(x)= \frac{1}{2}[\tanh(x/\sqrt{2} \xi)+1]$, both satisfying the B.C. and exhibiting approximately the same profile at the interface.} writing $f(x)=1-\epsilon(x)$, where $\epsilon(x) << 1$ and keeping only linear terms in $\epsilon(x)$:

\beq
\label{sol}
\epsilon(x)=e^{- \sqrt{2}x/ \xi}.
\eeq
With the boundary conditions (B.C.) $f(x \to \infty)=1$ and $f(x \to -\infty)=0$, the satisfactory solution is $f(x)=(1-\epsilon(x))\Theta(x)$. Since inside the superfluid region $\Delta$ does not vary, $A$ has to be identified with the BCS gap parameter. Namely, $A=\Delta_0$. Thus the order parameter $\Delta(x)$ takes the form
\beq
\label{sol2}
\Delta(x)= \Delta_0[1-e^{- \sqrt{2}x/ \xi}]\Theta(x).
\eeq
In Fig.~(\ref{gapprofile}) the gap profile is sketched as a function of $x$ at the interface between the normal and superfluid phases. The energy density associated with the interface {\it a la} GL is

\bea
\label{en1}
E(\Delta(x)) =\alpha \Delta(x)^2 +\frac{\beta}{2} \Delta(x)^4  -\gamma \Delta(x)  \nabla^2 \Delta(x).
\eea
From the equation above we see that deep inside the superfluid phase we have $E(\Delta(x \to \infty))=\frac{\alpha}{2}\Delta_0^2\equiv E_0$, and inside the normal phase, $E(\Delta(x \to -\infty))=0$. Since a stable interface requires (in order to fulfill mechanical equilibrium) equal energy densities on both the normal and superfluid sides, we must have that the energy inside the normal phase is also $E_0$. This is accomplished if we define a total energy density

\bea
\label{en2}
E_T(\Delta(x)) =E(\Delta(x)) + \Theta(-x) E_0,
\eea
The equation above clearly satisfies $E_T(x \to \infty)=E_T(x \to -\infty)=E_0$. Then, the surface (excess) energy density is defined to be:
\bea
\label{se}
\sigma =\int_{-\infty}^{\infty} \left[E_T(\Delta(x)) - E_0 \right] dx,
\eea
where the term subtracted in the equation above is the energy of the fully superfluid phase (or the fully normal state). If we multiply Eq.~(\ref{gl}) by $\Delta$, integrate with respect to $dx$ and substitute the resulting equation into the equation above, it is then easy to show that the (stationary) surface energy is given by:

\beq
\label{se2}
\sigma= \frac{\alpha \Delta_0^2}{2} \left\{  \int_{-\infty}^{\infty} [ \epsilon(x)^4 + \Theta(-x) -1] dx \right\}.
\eeq
In Eq.~(\ref{se2}) we have eliminated $\beta$ in favor of $\alpha$. The $x$ (numerical) integration yields

\beq 
\label{se5}
\sigma  = \left\{ -\frac{25}{24} \right\} \sqrt{2} \xi \frac{\alpha  \Delta_0^2}{2}.
\eeq
Before pursuing further, it is worth mentioning that the surface tension could be obtained by pure phenomenological considerations. It is given by the product of the typical energy density difference between the superfluid and normal phases and the correlation length, as we have obtained. Nevertheless, the microscopic derivation serves to clarify and validate the results and allows direct extension to other systems. The result in Eq.~(\ref{se5}) is for $\alpha<0$, which implies positive surface energy and could, {\it a priori}, favor the HGS phase relative to the MP state. However, $\alpha<0$ (or $\frac{1}{\bar{k}_F a}>-\frac{4}{\pi}$) characterizes a system in the strong coupling limit which is beyond the validity of the mean field BCS approximation. In this range at least the leading order $\bar{k}_Fa$ corrections to the pressure and chemical potential should be included~\cite{Carlson}. We define:

\beq
\label{se6}
\sigma^{+} \equiv \sigma \left(1/\bar{k}_F a >-4/\pi \right) =  \frac{25 \times \sqrt{2}}{48} |\alpha| \xi \Delta_0^2,
\eeq
where the true range of strong coupling in the equation above, such that the BCS part of the MP is a superfluid, is $-4/\pi < 1/\bar{k}_F a < 0$. As we will see below, couplings beyond this point (i.e., $1/\bar{k}_F a > 0$), correspond to weakly coupled molecules in BEC~\cite{Leggett}.

\begin{figure}[t]
\includegraphics[height=2.4in]{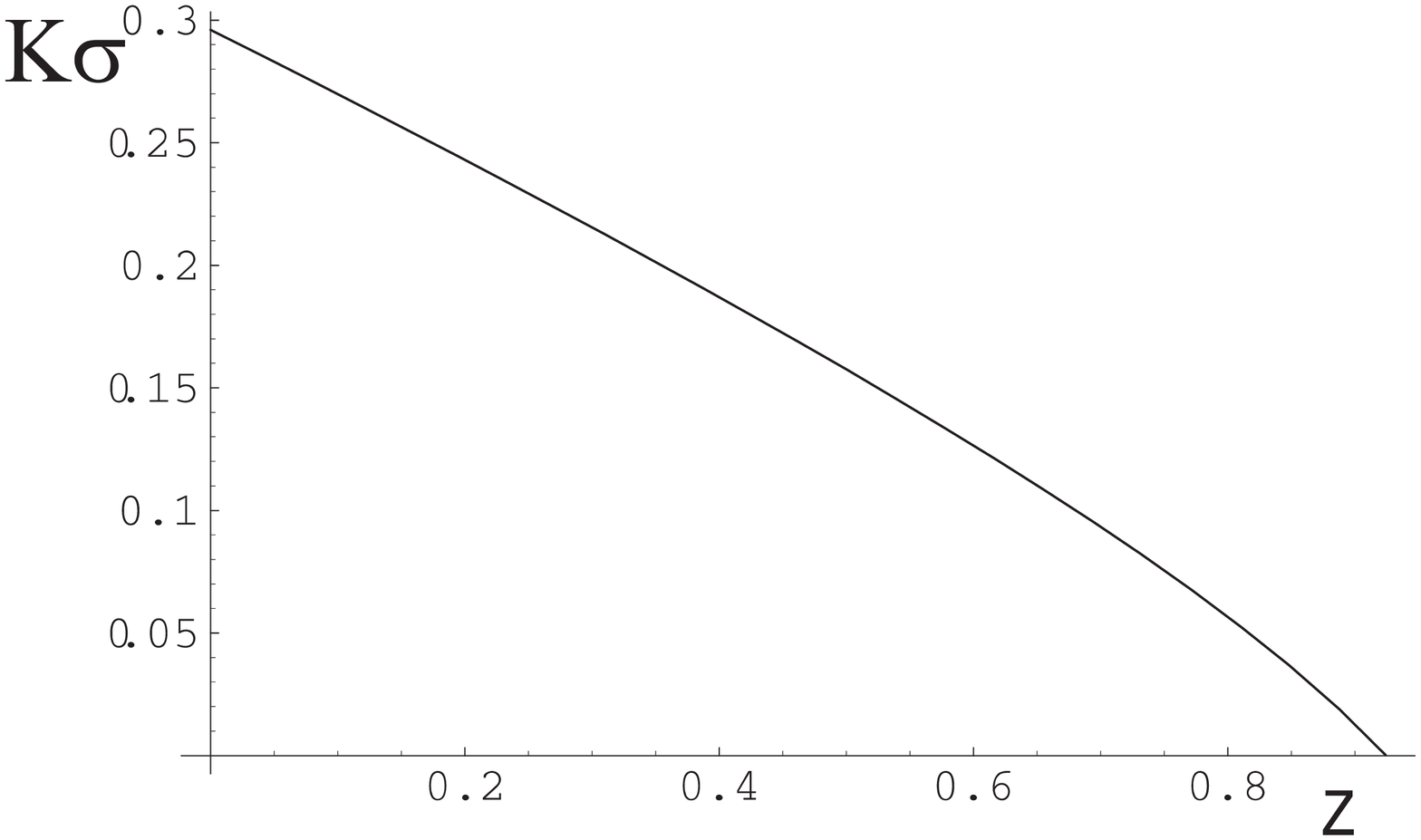}
\caption{\label{SurfaceEnergy}\textit{Surface energy between normal and superfluid phases in the BCS regime $ 1/\bar{k}_F a \leq -4/\pi$ (i.e., $a$ small and negative). We see that $\sigma^{\alpha \geq 0}$ is maximum for $z=4\delta \alpha /\beta^2=0$ or $1/\bar{k}_F a = -4/\pi$, and vanishes for $z=z_{max}$ or $1/\bar{k}_F a = -4/\pi(1+\frac{z_{max} \beta^2}{4 N_0 \delta}).$} }
\end{figure}

In the case where $\alpha > 0$ or, in the weak coupling limit $ 1/\bar{k}_F a < -4/\pi$ (i.e., $a$ small and negative), one would have to go to sixth order in the expansion in $\Delta$. Then the energy expression, Eq.~(\ref{en1}), is now

\bea
\label{enc}
E(\Delta(x,\delta)) =\alpha \Delta(x)^2 +\frac{\beta}{2} \Delta(x)^4 +\frac{\delta}{3} \Delta(x)^6 -\gamma \Delta(x)  \nabla^2 \Delta(x).
\eea
The energy will have a stable configuration i.e.,bounded from below, (for $\alpha \geq 0$) provided the signs of the expansion coefficients are $\beta<0$ and $\delta>0$, which in turn depend on the chemical potentials asymmetry. The space independent non-trivial equation for the minimum of Eq.~(\ref{enc}) is $\alpha-|\beta| \Delta_0^2+\delta\Delta_0^4=0$, whose solution is

\bea
\label{enc2}
\Delta_{0,\pm}^2=\frac{\beta}{2 \delta}\left[-1\pm \sqrt{1-4\alpha \delta /\beta^2} \right].
\eea
A real and positive solution also requires $\beta^2>4\alpha \delta$. In the point where $\alpha=0$, we have $\Delta_{0,+}^2=0$, corresponding to the normal solution, and $\Delta_{0,-}^2=|\beta|/\delta$ representing a superfluid solution. Thus the solution which gives the minimum for $\alpha\geq0$ is $\Delta_{0,-}^2=\frac{|\beta|}{2 \delta}\left[1 + \sqrt{1-4\alpha \delta /\beta^2} \right]$. Proceeding as before, the surface energy is written as

\bea
\label{seGeral1}
\sigma^{\alpha \geq 0} =\int_{-\infty}^{\infty} \left[E_T(\Delta(x))^{\alpha \geq 0} - E_0^{\alpha \geq 0} \right] dx,
\eea
To derive the expression above we have used the equation of motion $[\alpha + \beta \Delta(x)^2  + \delta\Delta(x)^4 - \gamma \nabla^2] \Delta(x)=0$, which yields

\bea
\label{seGeral2}
E_T(\Delta(x))^{\alpha \geq 0} = \frac{|\beta|}{2} \Delta(x)^4 -\frac{2\delta}{3} \Delta(x)^6+ \Theta(-x)E_0^{\alpha \geq 0},
\eea
where $E_0^{\alpha \geq 0}=\frac{2}{3}\alpha \Delta_{0,-}^2 -\frac{|\beta|}{6} \Delta_{0,-}^4 =\frac{|\beta|}{2} \Delta_{0,-}^4 \left(1-\frac{4}{3}\frac{\delta}{|\beta|} \Delta_{0,-}^2 \right)$, due to the minimum condition. If we define a real function as $\Delta(x)=\Delta_{0,-} g(x)$, the equation for $\Delta(x)$ is now

\bea
\label{seGeral3}
c_1 g(x) -c_2 g(x)^3 + c_3 g(x)^5 -c_4 \frac{d^2}{dx^2}g(x)=0,
\eea
where the coefficients $c_1$ to $c_4$ are all positive, and given by

\bea
\label{seGeral4}
c_1&=&\alpha;\\
\nonumber
c_2&=&|\beta| \Delta_{0,-}^2;\\
\nonumber
c_3&=&\delta \Delta_{0,-}^4;\\
\nonumber
c_4&=&\gamma.\\
\nonumber
\eea
We write $g(x)=1-\chi(x)$, with $\chi(x)<<1$, and get after keeping only linear terms in $\chi(x)$,

\bea
\label{seGeral5}
\frac{d^2}{dx^2}\chi(x)=C_1\chi(x)+C_2,
\eea
where $C_1=\frac{ 5c_3+c_1-3c_2}{c_4}$ and $C_2=\frac{c_2-c_1-c_3}{c_4}$. The equation above can be solved as $\chi(x)=y(x)-C_2/C_1$, which yields $y(x)=e^{-\sqrt{C_1} x}$. However, $C_2=0$ as a consequence of the minimum condition, and the final solution for the gap parameter satisfying the B.C. is

\bea
\label{seGeral6}
\Delta(x)=\Delta_{0,-}[1-e^{-\sqrt{C_1} x}]\Theta(x).
\eea
The behavior of $\Delta(x)/\Delta_{0,-}$ is the same as the one shown in Fig.\ref{gapprofile}. Then, Eq.~(\ref{seGeral1}) takes the form

\bea
\label{seGeral7}
\sigma^{\alpha \geq 0} &=&\int_{-\infty}^{\infty} \left[\frac{|\beta|}{2} \Delta(x)^4 -\frac{2\delta}{3} \Delta(x)^6+ \Theta(-x)E_0^{\alpha \geq 0} - E_0^{\alpha \geq 0} \right] dx\\
\nonumber
&=&\frac{|\beta|}{2} \Delta_{0,-}^4 \int_{-\infty}^{\infty} \left[g(x)^4 \left(1-\frac{4}{3}\frac{\delta}{|\beta|} \Delta_{0,-}^2 g(x)^2  \right) - \left(1-\frac{4}{3}\frac{\delta}{|\beta|} \Delta_{0,-}^2 \right) \right]\Theta(x) dx,
\eea
where we have used that $\Theta(-x)-1=-\Theta(x)$. We use the equation for $\Delta_{0,-}^2$ in the equation above to obtain

\bea
\label{seGeral8}
\sigma^{\alpha \geq 0} &=&\frac{9 |\beta|^3 \bar \xi}{32 \sqrt{2} \delta^2} g(z)^2  \int_{0}^{\infty} \left\{g(u)^4 \left[1- g(z) g(u)^2  \right] - \left[1-g(z) \right] \right\} du\\
\nonumber
&=& \frac{9 |\beta|^2 \sqrt{\gamma}}{16 \sqrt{2} \sqrt{\delta^3}} \frac{g(z)^2}{L(z)} I(z),
\eea
where $\frac{\bar \xi}{\sqrt{2}}=\frac{1}{\sqrt{C_1}}=\frac{1}{\sqrt{2}}\sqrt{\frac{\gamma}{   \delta \Delta_{0,-}^4-\alpha}}$, $g(u)=1-e^{-u}$, $g(z)=\frac{2}{3}[1+\sqrt{1-z}]$, with $z=\frac{4 \alpha \delta}{\beta^2}$, $0\leq z \leq z_{max}$, where $z_{max}\leq 1$ will be find such that $\sigma^{\alpha \geq 0}=0$. Then we read the limits for the interaction parameter from $z$ in order to have a real surface energy: $0 \leq \alpha \leq z_{max}\beta^2/4\delta$. In the second line of the equation above we have written $\sqrt{   \delta \Delta_{0,-}^4-\alpha}=\sqrt{9 \beta^2/8\delta~ g^2(z)-\beta^2z/4\delta }=|\beta|/(2\sqrt{\delta})~L(z)$, where $L(z)=\sqrt{(9/2)g^2(z)-z}$. Integrating in $u$ we find that $I(z)=-0.449+1.633\sqrt{1-z}$. Solving $I(z)=0$ we obtain $z_{max}=0.924$. For fixed $\delta \mu$, as is relevant to experiments, the parameters $\beta$, $\delta$ and $\gamma$ will also be fixed, and the only parameter of interest is $1/k_Fa$ encoded in $\alpha$. Then we plot in Fig.~\ref{SurfaceEnergy} the function $K \sigma^{\alpha \geq 0}$, where $K=\frac{ \sqrt{\delta^3}}{\beta^2 \sqrt{\gamma}}$ as a function of $\frac{9g(z)^2}{16 \sqrt{2}L(z)} I(z)$, for $z$ varying from $0$ to $z_{max}$. The behavior of $K \sigma^{\alpha \geq 0}$ shows that the surface tension goes to zero for $1/k_Fa <<-1$.

\begin{center}
B. $\mu<0$, $a>0$
\end{center}
For completeness we analyze now the case $a>0$ in the BEC regime. We focus only on the space-independent solution of Eq.~(\ref{gl}). After taking into account the normal energy, Eq.~(\ref{en1}) at the minimum reads:

\beq
\label{m}
E(\Delta)=E^N+\frac{\alpha}{2}\Delta^2,
\eeq
where $E^N=-\frac{4}{15} \frac{m^{3/2}}{\sqrt{2}\pi^2}(h-|\mu|)^{5/2}\Theta(h-|\mu|)$~\cite{Leo}. In the BEC regime, where $\mu <0$, $\alpha=\frac{m}{4 \pi a}- \frac{m \sqrt{2m|\mu|}}{4 \pi}$, ($\alpha=V_2$ in Ref.~\cite{Leo}). In the case $h < |\mu|$, $\alpha \approx \frac{m \bar{k}_F}{4 \pi} \left(\frac{1}{\bar{k}_F a} - 1 \right)$. The point where $\alpha=0$ gives $\frac{1}{ \bar{k}_F a} = 1$. Substituting $\bar{k}_F=\frac{k_F^a+k_F^b}{2}=\frac{\sqrt{2m \mu_a}+\sqrt{2m \mu_b}}{2}$ in this equation, we find:

\beq
\label{bound1}
\sqrt{\mu_a}+\sqrt{\mu_b}=\frac{2}{\sqrt{2m}~a}.
\eeq
Defining 
\bea
\label{bound2}
\mu_a=|\mu|+h,\\
\nonumber
\mu_b=|\mu|-h,
\eea
we get for $h < |\mu|$ up to order $h^2$

\beq
\label{bound3}
\mu(h)= - \frac{E_b}{4} -\frac{1}{2} \sqrt{\left(\frac{E_b}{2}\right)^2-h^2},
\eeq
where $E_b= \frac{1}{ m a^2}$. This is a well-known result for $h=0$ where, in this limit, tightly bound pairs with binding energy $E_b$, and nondegenerate fermions with a large and negative chemical potential are expected~\cite{Melo,Leo}. Eq.~(\ref{bound3}) serves, in this approximation, to define a critical value for the chemical potentials asymmetry, $h_c=\frac{\delta \mu_c}{2}=\frac{E_b}{2}$. Then, $0\leq h \leq h_c$.

The pair field $\Delta$ for $h < |\mu|$ in this moderate coupling ($ 1/\bar{k}_F a \approx 1$) regime, where the mean field is not applicable, has to be determined by the number equation,

\beq
\label{d}
n=-\frac{\partial E}{\partial \mu} =\frac{\Delta^2}{2} \frac{\partial \alpha}{\partial \mu}=\frac{m^2 \Delta^2}{8 \pi \bar{k}_F},
\eeq 
which gives,

\beq
\label{d2}
\Delta=\sqrt{n \frac{8 \pi \bar{k}_F}{m^2}}.
\eeq 
We plug in the equation above the total (free) density 

\beq
\label{d3}
n=n_a + n_b = \frac{ {k_F^a}^3 }{6 \pi^2} + \frac{ {k_F^b}^3 }{6 \pi^2} = \frac{ {\bar{k}_F}^3 }{6 \pi^2}g(h/|\mu|),
\eeq
where $g(h/|\mu|)\equiv (1+h/|\mu|)^{3/2}+(1-h/|\mu|)^{3/2}= 2 + \frac{3}{4}\left(\frac{h}{|\mu|}\right)^2+O(\left(h / |\mu|\right)^4)$, to find

\beq
\label{d4}
\Delta(h < |\mu|)=\sqrt{ \frac{16}{3\pi} g(h/|\mu|)}~|\mu|,
\eeq 
where $\mu$ is given by Eq.~(\ref{bound3}). From the equation above we see that the role of the chemical potential asymmetry is to lower the pairing field gap, as compared to the symmetric case ($h=0$). For $h > |\mu|$, $\alpha=\frac{m}{4 \pi a}- \frac{m \sqrt{2m|\mu|}}{4 \pi}=0$ yields

\beq
\label{d5}
\mu= - \frac{E_b}{2}.
\eeq 
The density now has the contribution of the normal energy:

\beq
\label{d6}
n=-\frac{\partial E^N}{\partial \mu}+\frac{\Delta^2}{2} \frac{\partial \alpha}{\partial \mu}=\frac{2}{3} \frac{m^{3/2}}{\sqrt{2}\pi^2}(h-|\mu|)^{3/2}+\frac{m^2 \Delta^2}{8 \pi \sqrt{2m|\mu|}}.
\eeq 
Using $n=\frac{{(2m|\mu|)}^{3/2}}{3 \pi^2}$, we finally obtain

\beq
\label{d7}
\Delta^2(h > |\mu|)= \frac{8  \sqrt{2m|\mu|}}{m^2}\left[\frac{{(2m|\mu|)}^{3/2}}{3 \pi}-\frac{2}{3} \frac{m^{3/2}}{\sqrt{2}\pi}(h-|\mu|)^{3/2} \right],
\eeq
where $\mu$ is given by Eq.~(\ref{d5}). The chemical potential asymmetry for $h > |\mu|$ also has the effect of lowering the pairing gap, as happens for $h < |\mu|$. The equation above sets the limit $h_{c1}<h \leq h_{c2}$ where $h_{c1}=\frac{E_b}{2}$, and $h_{c2}=(1+2^{2/3})\frac{E_b}{2}$.

The study of space dependent solutions of Eq.(\ref{gl}) for $a > 0$, as well as the consideration of a possible surface energy between a phase with dilute free fermionic atoms, and a phase with condensed molecules within the BEC regime~\cite{Leo,Sa}, needs further investigation and will be considered in future work.

In conclusion, we have obtained the surface energy that could arise in cold asymmetrical fermionic systems. The interaction limit, $-4/\pi < 1/\bar{k}_F a \leq 0$ (which implies $\alpha<0$, and consequently $\sigma>0$), shows that in the strong coupling regime (including the unitary limit i.e., infinite scattering length limit) a positive surface energy will emerge. We have shown that in the weak coupling BCS regimes, $ 1/\bar{k}_F a \leq -4/\pi$, where phase separation exists, it is found to be stable with no surface energy cost. In this regime, we found a lower limit for $1/\bar{k}_F a$, which is simply an artifact of the truncation of expansion of the energy at sixth order. However, the vanishing of the surface energy for small $\bar{k}_F a$ is an accurate and expected result. This is in complete accordance with the mean-field and Quantum Monte Carlo methods employed by Carlson and Reddy~\cite{Carlson}. Their results indicate that a HGS would exist only at very strong coupling, where $\Delta/\mu \gtrsim 1$. They expect that in this regime, the surface energy cost could help stabilize the HGS phase over the MP state. 

We also found the expressions which fix $\mu$ in the BEC regime for $h<|\mu|$ and $h>|\mu|$, in terms of the molecular binding energy $E_b$. We found that in both cases the role of the chemical potentials difference is to decrease the pairing gap size.

We hope that the surface energy we have obtained could also be used in the theoretical investigations of the BEC-BCS crossover in asymmetrical fermionic systems (since in this entire regime of coupling, more coexisting exotic phases have been predicted to occur~\cite{Dan}), and also in the studies of superfluid properties of neutron matter.
\\
\\
{\bf Note added:} After completion of this work, a preprint appeared on arxiv~\cite{Mueller}, where there is investigated the role of the surface tension between the normal and superfluid regions of a trapped fermion gas at unitarity ($k_Fa \to \infty$). They find in this regime that the surface tension is responsible for distortions in the atomic cloud in the boundary between these two regions. Although De Silva and Mueller use a different approach to obtain the domain wall energy at the superfluid-normal interface, they use the same expansion to calculate the coefficients $\alpha$ and $\gamma$.
\\
\begin{acknowledgments}
I thank the Nuclear Theory Group of the Lawrence Berkeley National Laboratory for their hospitality, where part of this work was done. I also thank P. Bedaque for enlightening conversations and for reading the manuscript, A.~L. Mota, B.~V. Schaeybroeck, G. Rupak and M.~Zwierlein for valuable discussions. This research was partially supported by FAPEMIG and CNPq/Brazil.

\end{acknowledgments}

\end{document}